\documentclass[prb,aps,twocolumn,showpacs]{revtex4}
\usepackage[dvips]{graphicx}
\usepackage{amsmath,amsfonts,amssymb,bm,ulem,color}

\begin{document}

\title{Collective dynamics of interacting Ising spins:
Exact results for the Bethe lattice}
\author{A. L. Burin}
\affiliation{Department of Chemistry, Tulane University, New Orleans LA, 70118}
\author{N. V. Prokofev}
\affiliation{Department of Physics, University of Massachusetts, Amherst, MA 01003}
\author{I. S. Tupitsyn}
\affiliation{Pacific Institute of Theoretical Physics, University of
British Columbia, Vancouver, BC V6T1Z1, Canada}
\date{\today}
\begin{abstract}
We study the low temperature dynamics in films made of molecular magnets, i. e. crystals
composed of molecules having large electronic spin $S$ in their ground state. 
The electronic spin dynamics is mediated by
coupling to a nuclear spin bath; this coupling allows transitions
for a small fraction of electronic spins between their two energy
minima, $S^{z}=\pm S$, under resonant conditions when the change of
the Zeeman energy in magnetic dipolar field of other electronic
spins is compensated by interaction with nuclear spins. Transitions
of resonant spins can result in opening or closing resonances in
their neighbors leading to the collective dynamics at sufficiently
large density $P_{0}$ of resonant spins. We formulate and solve the
equivalent dynamic percolation problem for the Bethe lattice (BL) of
spins interacting with $z$ neighbors and find that depending on the
density of resonant spins $P_{0}$ and the number of neighbors $z$
the system has either one ($2<z<6$) or two ($z\geq 6$) kinetic
transitions at $P_{0}=P_{c1} \approx e^{-1/3}/(3z)$ and
$P_{0}=P_{c2} \approx e^{-1}/z$. The former transition is continuous
and associated with the formation of an infinite cluster of coupled
resonant spins similarly to the static percolation transition
occurring at $P_{0}\approx 1/z$. The latter transition, $z>5$, is
discontinuous and associated with the instantaneous increase in the
density of resonant spins from the small value $\sim 1/z$ to near
unity. Experimental implications of our results are discussed.
\end{abstract}


\maketitle

\section{Introduction}

Percolation theory describes the flow in heterogeneous media. It is
successfully applied to a variety of physical, chemical, biological
and even social processes ranging from the hopping conductivity in
doped semiconductors \cite{ShklovskiiBook} to the evolution of large
genetic networks and self-organized criticality
\cite{Science1,RevModPhysGen}. For instance, the low temperature
conductivity of hopping insulators can be modeled by the equivalent
network of resistances replacing elementary electron hopping events
\cite{ShklovskiiBook}. The percolation theory finds the optimum set
of interconnected open channels characterized by resistances not
exceeding  a certain maximum and forming  an infinite cluster to
guide electrons through the sample.

Open percolation channels are usually treated as stationary in time.
It is relatively easy to study the static case numerically. Moreover
there exist analytical solutions including percolation on the Bethe
lattice (BL) and in low-dimensional systems
\cite{StaticCayley,Kirkpatrick,avalanch,Binder,2d}. In all cases the
percolation kinetic transition is found to be continuous. System
parameters including, for instance, diffusion coefficient,
correlation radius and average size of percolating clusters all show
scaling behavior near the critical point similarly to the one for
second order phase transitions.

The model of static percolation is often an approximation. Flowing
particles interact with each other and their motion can open or
close percolation channels leading to, for instance, conductivity
noise  \cite{ab1}. The significance of fluctuations for the
cooperative dynamics of quantum defects, spins in spin glasses,
protons in ionic conductors and electrons in hopping insulators was
pointed out  in Refs. \cite{KM1,BMK,SpinGlass,Kozub1,Mark,Efros}. In
a more refined analysis of percolation one should include the
possibility that the flows through open channels can open or close
other channels. These dynamic correlations can affect kinetic
transitions  in a fundamental way and lead to the cooperative
dynamics, e. g. percolation may occur at smaller density of open
channels and become discontinuous. In this paper we suggest a simple
model of dynamic percolation for the low-temperature kinetics of the
system of interacting magnetic molecules and obtain an exact
description of the kinetic transitions on BL. According to the
arguments of Refs. \cite{Levitov,Mirlin} the use of BL is more
justified in our case than for the static percolation because of the
similarity in the phase space structure for Bethe lattice and
many-body system of interacting spins. In both cases the size of the
phase space grows exponentially with the system size ($2^{N}$ for
$N$ spins and $z(z-1)^{N-1}$ for the number of nodes in BL with the
coordination number $z$ and $N$ shells of the tree).


%

In the present work we study the dynamic percolation
model on the Bethe lattice and then project our results on the electronic spin
dynamics in films made of magnetic molecules such as $Mn_{12}$, $Fe_{8}$, etc.
These molecules are usually composed of transition metals, forming the magnetic core, coupled to
various organic ligands \cite{BB1}. They possess large electronic
spin $S$ in the ground state (both $Mn_{12}$ and
$Fe_8$ molecules are characterized by the ground state ``central
spin" $S=10$). At low temperatures only the two lowest states of
each molecule, characterized by the spin projections $S^z_i=\pm S$
to the easy axis $z$, are occupied and each molecule can
be modeled as a two-level system with two states corresponding to
the Ising pseudo-spin $1/2$. The tunneling gap $2 \Delta_o$ between
the two lowest spin states in zero external transverse field is
tiny, $\sim 10^{-11} - 10^{-7} \; K$. It is much smaller than other
relevant parameters including the strength of the dipolar
interaction between the nearest-neighbor spins, $U_D \sim 0.1 \; K$,
and the half-width of the distributions of the nuclear biases, $E_o
\sim 10^{-2} - 10^{-3}$K. Under these conditions only
spins exposed to a total longitudinal bias (external field plus
internal dipolar demagnetization field) smaller than $E_o$ can
efficiently transfer between their two states due to energy exchange
with the nuclear spin bath \cite{PrStPRL98,Stamp2,Stamp1}. These
spins are the resonant, or ``open", ones. At the same time, since
$U_o \geq E_o$, each spin flip affects all its neighbors and may
create new resonances (or destroy existing ones, see Sec.
\ref{sec:spindyn}). 
The associated collective dynamics of spins is
studied within the dynamic percolation model in Secs.
\ref{sec:dynpercmod}, \ref{sec:dynpercsol} (see Fig.
\ref{fig:Resonance}).

 To be more specific, one may be interested in the correlation function
$C(t) = N^{-1} \sum_i \langle S^{z}_i(t)  S^{z}_i(0) \rangle $ where $N$ is the total
number of spins, and the average is taken over initial conditions, and evolution histories.
An interesting quantity to investigate in the infinite time limit would be the fraction of
spin which never change their magnetization $W_*$. This fraction can be easily seen to result
in the finite value of $C(t \to \infty )$ since dynamically frozen spins contribute unity to 
the sum above. The other question concerns spin diffusion and the possibility
to transfer energy and magnetization over large distances in which case one should be looking for
the onset of percolating clusters of mobile spins. 


The effect of dynamic and percolation transitions on spin
relaxation in films of magnetic molecules, based on examples of $Mn_{12}$ and Fe$_{8}$, 
is discussed in Sec. \ref{sec:Experiment}. Our goal here is to investigate the
case of completely demagnetized samples at relatively high temperatures
$k_B T > U_D$ to avoid the ``dipolar ordering" effects and related 
sample-geometry dependent demagnetization fields. 
In this limit the
resonant spins are nearly uniformly and randomly distributed in the system
bulk contrary to the case of strongly polarized samples where
resonant spins form spatially ordered structures, the ``resonant
surfaces", and initial conditions play an important role in the
system kinetics \cite{Stamp1}.

We predict the reduction of the relaxation rate by many
orders of magnitude in the vicinity of the transition point. At the
same time the abrupt change of the relaxation rate is smeared out by
its continuous dependence on the total longitudinal bias, which may
obscure the experimental observation of the transition point.


\section{Collective dynamics of interacting spins}
\label{sec:spindyn}

Since the model of spin dynamics is already formulated
in great detail in Refs. \cite{PrStPRL98,Stamp1,Stamp2}, here we
briefly outline its main features and then introduce its Bethe
Lattice version. As it was mentioned in the Introduction,
the weakness of spin tunneling in molecular magnets allows an accurate
representation of the system in terms of Ising pseudospins $1/2$ 
coupled to each other by the long-range magnetic dipolar interaction
$U_{ij}S_{i}^{z}S_{j}^{z}$ with
\begin{equation}
U_{ij} = U_{D} a^3 \frac{1-3n_{ij}^{2}}{R_{ij}^{3}}, \;
\bm{n}_{ij}=\frac{\bm{R}_{ij}}{R_{ij}}, \label{eq:magn_dip}
\end{equation}
where $U_{D}$ is the strength of the dipole-dipole
interaction of spins; $a$ is the lattice constant; and ${\bf
{R}_{ij}}$ is the vector connecting molecules $i$ and $j$. For the
sake of simplicity we limit our consideration to the case of $2D$
square lattice. In this model each spin ${\bf S}_i$ is subjected to
the ``molecular'' field of all other spins
\begin{equation}
U_{i}=\sum_{j}U_{ij}S_{j}^{z}
\label{eq:MolField}
\end{equation}
and the external magnetic field $\mu h$, where $\mu$ is the related
magnetic moment.

The spin dynamics is associated with the spin tunneling between
states $S^{z}=\pm 1/2$. The transition of each spin between these
two states requires them to acquire or release the longitudinal
energy Eq. (\ref{eq:magn_dip}) to some thermal bath because of the
energy conservation. The coupling of spins to phonons usually
responsible for low-temperature dynamics in dielectrics is extremely
weak and can be neglected. The relaxation takes place due to the
interaction  of the central electronic spins with the
nuclear spin bath \cite{PrStPRL98,Stamp2,Stamp1}. Since in the limit
$\Delta_o << E_o$ only spins inside the ``resonance window" $2 E_o$
are allowed to tunnel, the approximate constraint $\mid \mu h +
U_{i} \mid < E_{o}$ determines the subspace of ``resonant spins". In
this form the constraint is most suitable for the dynamic
percolation model studied below in Sections \ref{sec:dynpercmod} and
\ref{sec:dynpercsol}. More accurate dependence of the relaxation time on energy 
is considered in Section \ref{sec:Experiment}.

\begin{figure}[ht]
\vspace{-2.8cm}
\includegraphics[width=8cm]{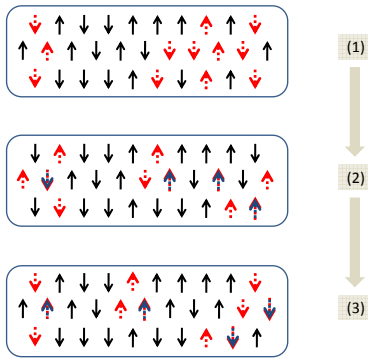}
\vspace{-3cm} \caption{ Evolution of ``percolating" (or ``open")
spins affecting their neighbors in $2D$ lattice. When some of the
open spins (shown by dashed red arrows) make transitions (shadowed
arrows for steps $(2)$ and $(3)$) they switch the state of the
neighboring spins between ``open" and ``close". }
\label{fig:Resonance}
\end{figure}

The only spins that are allowed to tunnel in any given
configuration are the resonant ones. Transitions of resonant spins
can change the status of their neighbors from non-resonant to
resonant and vice versa by changing their local bias from being
larger than $E_o$ to being smaller (see Fig. \ref{fig:Resonance}.
At the same time in the new configuration the transitions of
neighbors, diffused into the resonance window, can push the former
resonant spins out of resonance. Depending on the density of
resonant spins $P_{0} \sim E_o/U_{D}$ there exist two possibilities.
If $P_{0}$ is very small (i.e., the distance between the
resonant spins is large), the transitions of resonant spins can
essentially affect only their local environment and spins in the
sample can be separated into a small ``mobile'' and large
``immobile'' groups. The mobile group consists of percolating
spins, capable of making transitions in the course of system
evolution, while the immobile group consists of non-percolating
spins which cannot flip in spite of the field fluctuations caused by
 mobile spins. If the density $P_{0}$ is large, the
resonant spin transitions create or destroy resonances around them
leading to the collective dynamics involving practically all spins
after sufficiently long time.


In what follows we investigate the kinetic transition
between these two regimes. The study of realistic $2D$ lattice of
spins, coupled by the dipolar interaction, can not be done exactly
analytically and requires numerical simulations. However, we can
solve a similar problem on the Bethe lattice with random
interactions between the neighbors. We assume that spins $1/2$
occupy all sites of the Bethe lattice and each spin interacts with
all its $z$ neighbors.  The interaction constant $U_{ij}$ between
two neighboring spins is assumed to be the Gaussian random variable
with the zero average and dispersion $U_D$. This assumption is
approximate - dipolar interactions in a real system involve not only
nearest neighbors but also spins separated from the given one by
more than  one lattice constant. The long range nature
of dipolar interactions can not be neglected in three-dimensions,
but in the two-dimensional system the lattice sum of
$1/r^3$ is finite and the nearest neighbor approximation 
is justified. It is clear that the model we
study and the real problem differ from each other; however they can
belong to the same universality class because the phase-space
structure for interacting Ising spins is similar to that for the
Bethe lattice. The phase space grows exponentially with the number
$N$ of spins as in many-body problem \cite{Levitov,Mirlin}, which
permits us to neglect ``close loops'' for different evolution paths
of the system in its phase space thus making it similar to the Bethe
lattice.

As mentioned above, we restrict our considerations to the
 limit of low temperatures, where each molecule can
still be modeled as a two-level system, but we assume $k_{B} T >
U_{D}$ to avoid the dipolar ordering and, consequently, any
long-range statistical correlations between spins. Indeed, such
correlations should affect the kinetic transition. This assumption is not 
restrictive since, for instance, in $Fe_8$ and
$Mn_{12}$ the regime of pure ground-state tunneling can be reached
at temperatures $T \lesssim 0.4 \; K$ and $T \lesssim 0.8 \; K$
respectively which is larger than $U_D$ in both systems
\cite{TupitsynBarbara}. We also consider only completely depolarized
system here.

\section{Dynamic percolation problem on Bethe lattice}
\label{sec:dynpercmod}

In the Bethe lattice the spin is resonant (open) if its 
local bias (or Zeeman splitting) $U_{i}=\sum_{j}U_{ij}S_{j}$, Eq.
(\ref{eq:MolField}), is smaller than $E_{o}$. Since the dipolar
fields distribution in the depolarized limit is similar to the Gaussian one (see
Fig.\ref{fig:DoS} below), the fraction of resonant spins (i.e., the
probability that the spin is open) is determined by $P_{0} \approx
2 E_{o} /(U_{D} \sqrt{2 \pi z}) \ll 1$. We expect (and this will be
confirmed by the final result) that the collective dynamics can take
place if each spin has approximately one resonant neighbor (cf.
Ref.\cite{Binder}), which leads to the estimate for the transition
point as $P_{0c}\sim 1/z$. This happens at $E_o \sim
U_{D}\sqrt{\pi/(2z)}$. Near the transition point one can
approximately ignore correlations between different configurations
of neighbors corresponding to open states of the given spin. Indeed,
if the spin is open in the given configuration the probability that
it will be open again after $k<z$ turns of neighbors can be
estimated as $P_{k} \sim 2E_{o}/(U_{D} \sqrt{2\pi k})$. The number
of neighbor transitions $Z$ needed to bring the given spin back to
the open state can be estimated setting $\sum_{k=1}^{Z}P_{k} \approx
1$. One can see that at $E_o\sim U_{D}\sqrt{\pi/(2z)}$ the next
opening of the given spin can be expected, on average,  
after turns of around $z/2$ neighbors  so the new resonant environment
is fully different from the previous one.

Under above assumptions one can formulate the following model of
dynamic percolation on the BL, see Fig. \ref{fig:Resonance}.
Consider a lattice occupied by spins $1/2$ having random projection
$S^{z}=\pm 1/2$. Assume that in any given configuration of spins the
rules for deciding about the spin states are the same: the
probability that a given spin is open (i.e., ``dynamic") in each of
the $2^z$ configurations of neighbors is $P_0$. In other words, the
dynamics of the model is fixed by having a table for each spin with
$2^z$ open and close states (the fraction of closed states is
$W_{0}=1-P_{0}$). Correlations between open states of neighbors in
different configurations are ignored. Only open spins are allowed to
change their states. As time evolves, all open spins have equal
chances to make a transition. The spin overturn affects all its $z$
neighbors in a way that they have to change ``status" according to
their tables. Our main goal is to study the cooperative dynamics of
spins, characterized by the fraction $P_{*}=1-W_*$, which
were involved into dynamics at some stage (if the spin ever flips in
the course of evolution we call it ``percolating") and to find out
when an infinite percolating cluster is formed by those percolating
spins.


The problem under consideration has common features and differences
with the bootstrap percolation problem \cite{bp1,bp2} used as a
mathematically idealized model for such phenomena as nucleation and
growth applied in the study of crack formation, magnetic alloys,
hydrogen mixtures, and computer storage arrays.  It has been
extensively studied both for the $2D$ lattice and BL. The bootstrap
percolation problem can be formulated in terms of spins $1/2$ placed
in all lattice sites. The spin is open and allowed to flip if the
number of neighbors with $S^{z}=+1/2$ exceeds the predefined number
$0 < n \leq z$. The cooperative dynamics is determined by the
initial density $P_{0}$ of spins with positive projections and in
the most investigated case of $n=2$ this dynamics vanishes in a
discontinuous manner at sufficiently small $P_{0}$; for the $2D$
lattice the threshold density $P_{0c}$ approaches zero with
increasing the system size. In spite of similarities, our model
differs qualitatively because the bootstrap percolation results in
the irreversible evolution of spin configuration towards the
increase in the density of spins with positive projection, while in
our case the density of open spins practically does not fluctuate
and we study a reversible equilibrium dynamics. One consequence of
this difference is vanishing $P_{0c}$ in the bootstrap
percolation problem in the $2D$ lattice while in our case there is
no cooperative dynamics at sufficiently small density of open spins
$P_{0}<2^{-z}/z$ ($z=4$ in $2D$) because non-percolating spins form
an infinite cluster blocking such dynamics, cf. Ref.
\cite{ShklovskiiBook}.

The exact description of the dynamic percolation for BL can be
obtained similarly to Refs. \cite{StaticCayley,bp1} (see Fig.
\ref{fig:neighb_sp}). First, we calculate the density $P_{*}$ of
percolating spins. Note that percolating spins can be in the close
state for some time. Thus percolating spins include open spins and
all other spins which can enter an open state at some point in time.
In what follows we show that the density $P_{*}$ can undergo a sharp
raise (see Fig. \ref{fig:InfResDens}) with increasing the density of
open spins $P_{0}$ above some critical value $P_{c2}$. This
discontinuous transition happens after the formation of the infinite
percolating cluster at $P_0=P_{c1}$ which promotes the cooperative
dynamics.

\section{Kinetics transitions in Bethe lattice}
\label{sec:dynpercsol}

\begin{figure}[ht]
\vspace{-2.4cm}
\includegraphics[width=8cm]{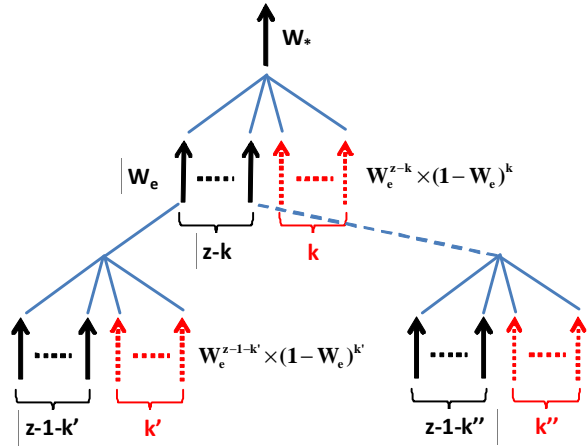}
\vspace{-2.4cm} \caption{ Configurations of neighboring spins in the
Bethe lattice and their probabilities. Red dotted arrows show percolating spins and black solid arrows  show non-percolating spins. Blue lines connect neighboring spins.}
\label{fig:neighb_sp}
\end{figure}

Consider the probability $W_{*}=1-P_{*}$ that a given spin is
non-percolating, i.e. it is never involved into dynamics despite
some of its $z$ neighbors making transitions on BL, see Fig.
\ref{fig:neighb_sp}. This probability depends on the states of all
its neighbors, which can be treated as un-correlated due to the
specific properties of BL. Each neighbor of the selected
non-percolating spin is characterized by the conditional probability
$W_{e}>W_{*}$ that it is also not percolating (see Fig.
\ref{fig:neighb_sp}). All neighbors are in identical situation by
construction.

The probability $W_{*}$ that the spin is not percolating can be
determined by considering different local environments distinguished
by the number of percolating neighbors $k=0, 1, ...z$, see Fig.
\ref{fig:neighb_sp}. There are $z!/(k!(z-k)!)$ independent ways to
be surrounded by $k$ percolating spins. In each of them the selected
spin will experience all $2^{k}$ possible states in the course of
evolution (assuming that percolating spins flip an unlimited number
of times; this is definitely true before the infinite cluster if
formed) and in each state it has to remain closed. The corresponding
probability is $(W_{0})^{2^{k}}$ (recall that $W_{0}=1-P_{0}$).
Summing up all the probabilities for different $k$ we get
\begin{equation}
W_{*}=\sum_{k=0}^{z}\frac{z!(1-W_{e})^{k}W_{e}^{z-k}}{k!(z-k)!}W_{0}^{2^{k}}.
\label{eq:prob}
\end{equation}
The probability $W_{e}$ for neighbors of the given non-percolating
spin to be non-percolating as well is defined similarly to Eq.
(\ref{eq:prob}) (see Fig. \ref{fig:neighb_sp}). One should consider
only the remaining $z-1$ neighbors characterized by the same
probability $W_{e}$ to be non-percolating. Accordingly we obtain the
self-consistent equation for the probability $W_{e}$ in the form
\begin{equation}
W_{e}=\sum_{k=0}^{z-1}\frac{(z-1)!(1-W_{e})^{k}W_{e}^{z-k-1}}{k!(z-k-1)!}W_{0}^{2^{k}}.
\label{eq:self_cons_prob}
\end{equation}

\begin{figure}[ht]
\vspace{-2.4cm}
\includegraphics[width=8cm]{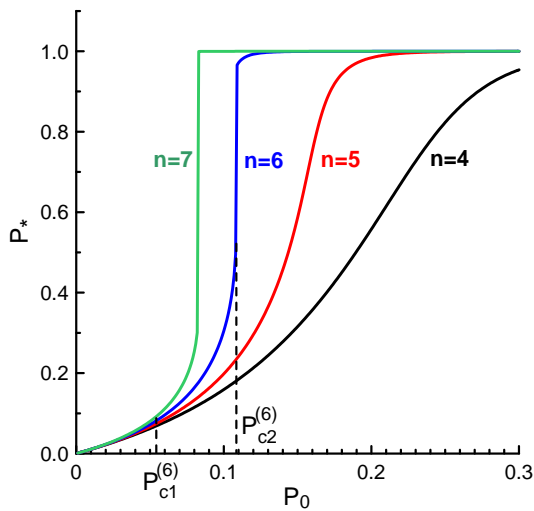}
\vspace{-1.8cm}
\caption{Dependence of the percolating sites density
$P_{*}$ on the density of open spins $P_{0}$ for $z=4$, $5$, $6$ and
$7$ neighbors. There is a discontinuous change in $P_{*}$ for $z>5$
where its density instantaneously jumps to $\approx 1$. Transition
points $P_{c1}\approx 0.0546$, $P_{c2}\approx 0.1085$ are shown for
$z=6$.} \label{fig:InfResDens}
\end{figure}

In Fig. \ref{fig:InfResDens} the solution of Eqs. (\ref{eq:prob}),
(\ref{eq:self_cons_prob}) for the density of percolating spins
$P_{*}=1-W_{*}$ is shown for BL with $z=4$, $5$, $6$ and $7$
neighbors. One can see that for $z=6$ and $z=7$ there exists the
discontinuous transition  at $P_{0}=P_{c2}$ (label $P_{c1}$ is
reserved for the formation of the infinite percolating cluster, see
below) where the density of percolating spins $P_{*}$ jumps from the
small value $P_{*} \sim 1/z$ almost to unity  $P_{*}
\sim 1-(W_{0})^{2^{z}}$. At $P_{0}>P_{c2}$ the vast majority of
spins belong to the infinite percolating cluster.

The physics behind the jump of the percolating spin density at the
transition point $P_{c2}$ can be illustrated in the limit of large
number of neighbors $z\gg 1$ when Eq. (\ref{eq:self_cons_prob}) can
be simplified by taking advantage of $P_{e}, P_{0}\ll 1$ and
expanding $(1-P_{0})^{2^{k}}\approx 1-2^{k}P_{0}$ (recall that
$P_{e}=1-W_{e}$)
\begin{equation}
P_{e}\approx P_{0}(1+P_{e})^{z-1}\approx P_{0}\exp(zP_{e}).
\label{eq:1/nlim}
\end{equation}
This equation has a solution only for $P_{0} <P_{c2}\approx
e^{-1}/z$, which agrees with the numerical solutions for $P_{c2}$ at
$z\geq 7$. In our opinion the discontinuous transition is caused by
an avalanche-type growth of the percolating site number in the
vicinity of other percolating sites. For instance, if the density of
percolating sites  $P_{*}$ reaches the value $1/z$ then
approximately each  site has one open neighbor. Its probability to
become percolating increases by the factor of $2$ which leads to the
formation of new open sites. This process leads to the jump in the
density of percolating spins to near unity.

The discontinuous transition was found in the bootstrap percolation
problem on Bethe lattice \cite{bp1}, but at effective parameter
$P_{0}=1/(z-1)^2$. It is interesting that in spite of the absence of
kinetic transition in the bootstrap percolation problem in the
infinite $2D$ lattice, the transition in the finite system is
essentially discontinuous \cite{bp2}. This gives us some hope that
the discontinuous dynamic percolation transition may take place in
our problem applied to realistic lattices of finite dimensions
($d>1$) similarly to a Bethe lattice.

The cooperative dynamics in the ensemble of interacting spins arises
in the presence of the infinite cluster of percolating sites. Such
cluster obviously exists at $P_{0}>P_{c2}$, but it may be formed at
smaller density $P_{0}$ of open sites. Indeed, the probability to
have a percolating site near another percolating site is about a
factor of $2$ larger than in the absence of percolating neighbors
because of the doubling in the number of explored configurations.
The corresponding critical point $P_{0}=P_{c1}$ where the infinite
cluster of percolating spins is formed can also be found exactly. It
can be shown (see Appendix) that the formation of the infinite
cluster is determined by the equation
\begin{eqnarray}
(z-1)(1+F_0-2F_1) =  \;\;\;\;\;\;\;\;\;\;\;\;\;\;\; \nonumber \\
1+(z-1)^2[ F_2(1-F_0)+F_1^2+F_0-2F_1 ] \;,
\label{eq:criterion_inf}
\end{eqnarray}
where $F_m$, see Eq.~(\ref{FN}) in the Appendix, is the probability
of finding a non-percolating spin surrounded by $z$ neighbors where
$m=0,1,2$ of them are \textit {definitely} percolating, $2-m$
neighbors are \textit{definitely} non-percolating and the remaining
group of $z-2$ neighbors can contain both percolating and
non-percolating spins. For small $P_0$ we have $F_m \approx 1-2^m
P_0\cdot(m(1-W_e)+W_e)^{z-2}$. The numerical analysis of the
transition point $P_{c1}$ corresponding to the formation of an
infinite percolating cluster confirms that  $P_{c1} < P_{c2}$. In
the $z\gg 1$ limit one finds that the infinite cluster is formed at
$P_{0}>P_{c1}\approx e^{-1/3}/(3z) <P_{c2}$ in agreement with the
numerical solution of Eq. (\ref{eq:criterion_inf}).

\section{Discussion. How can the kinetic transition be observed in
molecular magnets?} \label{sec:Experiment}

\begin{figure}[ht]
\vspace{-2.2cm}
\includegraphics[width=8cm]{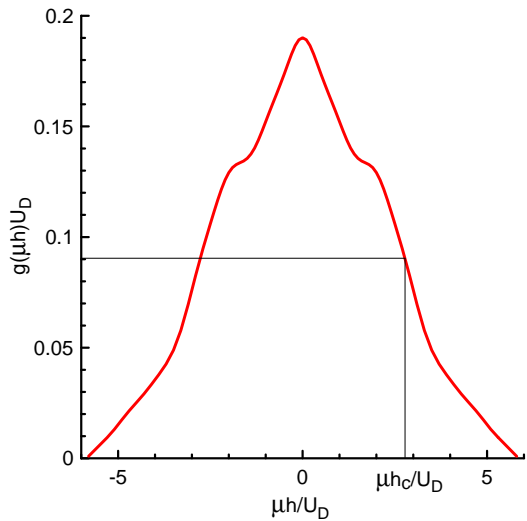}
\vspace{-1.8cm} \caption{Rescaled probability density for the spin
energy splitting on a model square lattice of magnetic
molecules. The predicted transition point is shown by solid lines.}
\label{fig:DoS}
\end{figure}

Below we consider the kinetic transition in a 2D square lattice
 of spins
representing 
magnetic molecules using the results obtained for the
Bethe lattice. This consideration should be applicable to the
recently synthesized two-dimensional crystals of 
Mn$_{12}$ molecules \cite{Andrea}. Assume that 
the easy axis is perpendicular to the sample plane.  Since only spins inside the resonance
window can change their state, the density of resonant spins in the
longitudinal magnetic field $h$ can be estimated as
\begin{equation}
P_{0} \approx 2 g(\mu h) E_o,
\label{eq:resprob}
\end{equation}
where $g(E)$ is the probability density for the Zeeman splitting $E$
(longitudinal bias) originated from the spin-spin
interaction (see Fig. \ref{fig:DoS}) and $\mu$ is the magnetic
moment of the molecule.

To study the kinetic transition using the previous results for the
Bethe lattice one has to introduce the number of neighbors parameter
$z$. We assume that the number of neighboring spins can be estimated
as the number of spins whose resonance can be affected by the
transition of the given spin. For the crude estimate we count
neighboring spins as those coupled to the given spin by the
interaction exceeding the width of the resonant window $2E_o$. In
the limit $E_o \ll U_D$ one can estimate this number in a
quasi-continuum approach as
\begin{eqnarray}
z \approx \pi (U_D/2E_o)^{2/3}.
\label{eq:numres}
\end{eqnarray}
According to the solution for the Bethe lattice (see Fig.
\ref{fig:InfResDens}) the sharp change of transition rate for almost
all spins should take place near the point $P_{0}=P_{c2}$. Assuming
$z \gg 1$  the crude estimate for the transition point can be made
using the approximate relationship $P_{c2}\approx e^{-1}/z$ (see
Sec. \ref{sec:dynpercsol}). Accordingly we get
\begin{equation}
E_{c} \approx \frac{U_{D}}{2(\pi e g(\mu h)U_{D})^{3}}.
\label{eq:2Dans}
\end{equation}
The distribution of the dipolar bias energies $g(E)$ in a sample
can be easily calculated, see Fig. \ref{fig:DoS}. The minimum threshold is
realized at zero field (maximum density of resonances) where $g(0) \approx 0.19/U_{D}$
so that we get $E_{c2}(0) \approx 0.12 U_{D}$.

Using the above analysis of the kinetic transition one can
approximately characterize the spin relaxation for different
molecular magnets. In $Fe_{8}$ one has $U_{D} \approx 130 \;$ mK and
$E_{o}\approx 6 \;$ mK \cite{StTuPRB04} so even in the absence of
the external field the system is in the localization regime. The
situation is different in $Mn_{12}$, where $U_D \approx 70 \;$ mK
and $E_{o} \approx 80 \;$ mK \cite{WWEPL99} and the collective spin
dynamics exists at zero magnetic field according to the theory, Eq.
(\ref{eq:2Dans}). However, application of the external longitudinal
magnetic field can reduce the resonance probability and result in
the localization. Using Eq. (\ref{eq:2Dans}) one can
estimate the value of the external field corresponding to the
transition point from 
\begin{equation}
g(\mu h_{c})U_{D} \approx \frac{1}{\pi e} \left (\frac{U_D}{2E_{o}} \right)^{1/3}
\approx 0.09
\label{eq:h2Dans}
\end{equation}
and the transition in $Mn_{12}$ takes place at the
field $h_{c} \approx 2.8 U_D / \mu \sim 0.2 \; T$ (see
Fig.\ref{fig:DoS}). One can also expect that the transition rate
should decrease near the transition point.

\begin{figure}[ht]
\vspace{-2.2cm}
\includegraphics[width=8cm]{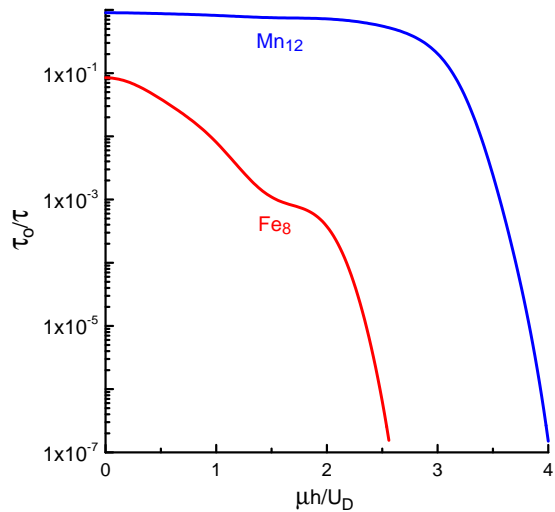}
\vspace{-1.8cm}
\caption{Dependence of the relaxation rate $\tau^{-1}$ on the magnetic field
near the transition point calculated using the exponential approach
Eq. (\ref{eq:rate_higEans}). $\tau_o$ is the characteristic relaxation time.}
\label{fig:Rel}
\end{figure}

Although theory predicts the absence of cooperative dynamics at
$h>h_{c}$ the reality is more complicated because spins which do not
belong to the transition window yet have small but finite transition
rate. Indeed, spins with the Zeeman energy $E>E_o$ can make
transitions; however their transition rate is becoming exponentially
small \cite{PrStPRL98,Stamp2,Stamp1}
\begin{equation}
\tau^{-1} \propto \exp(-\mid E\mid/E_{o}),
\label{eq:rate_higE}
\end{equation}
because such transition require simultaneous flips of large number
of nuclear spins. Then, we predict the exponentially small
transition rate which can be estimated following the standard
percolation theory approach \cite{ShklovskiiBook} as the rate Eq.
(\ref{eq:rate_higE}) at the Zeeman splitting $E$ equal to the
threshold value $E_{c}$, Eq. (\ref{eq:2Dans})
\begin{equation}
\tau^{-1} \propto \exp(-E_{c}/E_{o})\approx \exp(-(g(\mu h_{c})/g(\mu h))^3).
\label{eq:rate_higEans}
\end{equation}
This dependence is illustrated in Fig. \ref{fig:Rel} to show the
expected strong change in the rate of spin relaxation at large
fields. Unfortunately, the dynamics slowing down takes place at
fields $h \approx  2 h_{c}$, where the density of resonant spins is
already very small (see Fig. \ref{fig:DoS}) and the predicted
reduction of the relaxation rate by many orders of magnitude is more
difficult to study. The dynamic transition itself becomes a crossover 
because the transition rate is a continuous non-vanishing function 
of energy, Eq. (\ref{eq:rate_higE}). 

In addition to the application of the longitudinal magnetic field one can also affect the transition by 
diluting Fe$_8$ magnets. The reduction of the concentration of Fe$_{8}$ molecules to $x\ll 1$ will amount for changing $U_D/E_0$ ratio, and thus provide a knob for determining the transition point experimentally. Assuming that the effective constant for magnetic dipolar interaction scales proportionally to the $x^{3/2}$ one can estimate that at $x\leq 0.25$ the collective dynamics in $Fe_{8}$ films will take place.

\section{Conclusion}

We considered the dynamic percolation problem on the Bethe lattice
of spins $1/2$. Dynamics was introduced through the open spins
capable to change their states and affect the status of neighboring
spins switching them between open and closed states. The problem was
solved exactly. We found two kinetic transitions including the
continuous transition associated with the formation of the infinite
percolating cluster and the discontinuous transition associated with
the avalanche-type growth in the number of percolating sites. This
model approximately describes the low-temperature dynamics of
molecular magnets stimulated by their interaction with the nuclear
spin bath. In this model open spins are those having the small
Zeeman splitting compared to their hyperfine interaction. The sharp
kinetic transition is predicted for the $2D$ lattice of magnetic
molecules, however the abrupt change of the relaxation rate is smeared out by
its continuous dependence on the total longitudinal bias Eq. (\ref{eq:rate_higE}), which may
obscure the experimental observation of the transition point.

We acknowledge valuable discussions with Philip
Stamp and the support by the Pacific Institute of Theoretical Physics. 
The work of AB is supported by the Tulane Research and Enhancement
Fund. IT thanks the Tulane Research and Enhancement
Fund for supporting his visit to the Tulane University.

\section{Appendix}

 Consider the probability $P(N)$ that two spins separated by $N$
sites are connected through percolating sites. It is clear that if
this probability decreases with $N$ slower than the inverse number
of paths of length $N$ starting at the given site, $(z-1)^{-N}$,
then the infinite cluster of percolating sites is formed. This
criterion is also applicable to the static percolation problem on BL
where the connection probability is given by $P_{0}^{N}$.
Accordingly, the infinite cluster is formed at $P_{0}\geq 1/(z-1)$
in full agreement with the exact solution \cite{Binder}.


Each site of the line connecting spins $0$ and $N+1$ has $z-2$
neighbors which do not belong to the line. It is convenient for the
rest of the discussion to introduce a short notation for the
probability of finding a non-percolating spin (or $n$-spin) when
$m=0,1,2$ of them are \textit {definitely} percolating, $2-m$
neighbors are \textit{definitely} non-percolating and the remaining
group of $z-2$ neighbors can contain both percolating and
non-percolating spins.
\begin{equation}
F_m=\sum_{k=0}^{z-2}\frac{(z-2)!(1-W_{e})^{k}W_{e}^{z-k-2}}{k!(z-k-2)!}W_{0}^{2^{k+m}}
\;, \label{FN}
\end{equation}
cf. Eqs.~(\ref{eq:prob}) and (\ref{eq:self_cons_prob}). Indeed,
$F_m$ are all we need to know to calculate the probability that a
given line spin is non-percolating when the states of its line
neighbors are known. Since $W_e$ is known as a function of $W_0$, we
can concentrate on properties of the line spins alone. Note, that
this decomposition of probabilities Eq. (\ref{FN}) is possible due
to the Bethe lattice factorization around non-percolating spins. In
other words different realizations for spins belonging to branches
of BL separated by n-spins are independent of each other. Since the
states of the end spins ($0$ and $N+1$) can not change scaling
properties of the long line connecting them we select them to be
n-spins to simplify the derivation of $P(N)$. We then consider
$P(N)$ as complementary to the probability of having an arbitrary
line decomposition in terms of $p$- and $n$-spins with at least one
$n$-spin among $N>0$, i.e.
\begin{equation}
1=\sum_{\{n\}} \rho_{n,p} (\{n\}) \;,
\label{PN}
\end{equation}
where in each sequence $\{n\}$ all $N+2$ spins are divided into
$2s+1$ segments of alternating $n$- and $p$-spins as $\{n\}=(n_1,
p_1,n_2 \dots p_{s},n_{s+1})$ with $n_i, p_i \geq 1$ ($n_i$ is used
for n-spins and  $p_i$ for p-spins). The probability of a particular
sequence is denoted as $\rho_{n,p} [n_1, p_1, \dots , p_{s},n_{s+1}
]$. Clearly, $n_1+ p_1+ \dots + p_s + n_{s+1} = N+2$.

Due to factorization provided by n-spins (no information can be
exchanged between the neighbors of non-percolating spins and thus
all branches of BL around them are statistically independent) we
have
\begin{eqnarray}
&\rho_{n,p}& =F_0^{N} \;, \;\;\; (\mbox{for }\;s=0\;\mbox{or }\;n_1=N+2)\;,
\\
&\rho_{n,p}& =  \rho_{e} (n_1) \rho_{e} (n_{s+1}) \prod_{j=2}^{s-1}
\rho_n(n_j) \prod_{j=1}^{s} P(p_j)\;,\; (\mbox{for }\;s>0) \nonumber
\end{eqnarray}
Here $ \rho_n (m)$ is the probability of having a cluster of $m$
n-spins in a row, and, $\rho_{e} (m)$ is a similar quantity for the
first and last groups of n-spins. These probabilities can be
immediately computed by counting how many n-spins have $m$ p-spins
as their neighbors (recall that the end spins ($0$ and $N+1$) are
assumed to be non-percolating)
\begin{eqnarray}
 \rho_n (1)   &=  F_2\;,\;\;\;  \rho_n (m>1) = F_1^2F_0^{m-2}\;, \nonumber \\
 \rho_{e}(1)  &=  1\;,\;\;\; \rho_{e} (m>1)=F_1F_0^{m-2}\;.
\end{eqnarray}
The self-consistent equation (\ref{PN}) for $P(N)$ is solved using
generating functional $P(x)=\sum_{N=1}^{\infty} x^NP(N)$ (and
similarly defined $\rho_e(x)=\sum_{N=1}^{\infty} x^N\rho_{e}(N)$ and
$\rho_n(x)=\sum_{N=1}^{\infty} x^N\rho_{n}(N)$) which transforms
(\ref{PN}) into
\begin{equation}
\frac{x}{1-x}=\frac{F_0x}{1-F_0x} + \frac{\rho_e^2(x) P(x)}{x^2}
\sum_{k=0}^{\infty} [\rho_n(x)P(x)]^k  \;,
\label{PN2}
\end{equation}
with $\rho_e(x)= x + F_1x^2/(1-F_0x)$ and $\rho_n(x)= F_2x + F_1^2x^2/(1-F_0x)$.
After elementary algebra we find
\begin{equation}
P(x)=\frac{x^3(1-F_0)}{(1-x)(1-F_0x)\rho_e^2(x)+x^3(1-F_0)\rho_n(x)}\;,
\label{PN3}
\end{equation}
which can be further simplified to
\[
 P(x)=\frac{x(1-F_0)}{(1-x)(1-F_0x+2F_1x))+x^2[F_1^2+F_2(1-F_0)]} \;.
\]

The infinite cluster is formed if $P(N)$ decreases with $N$ slower
than $1/(z-1)^N)$. This means that the threshold is determined by
the divergence of $P(x)$ at $x=(z-1)$ which is only possible if the
denominator in Eq.~(\ref{PN3}) is zero. Thus the formation of the
infinite cluster is determined by the equation
\begin{eqnarray}
(z-1)(1+F_0-2F_1) =  \;\;\;\;\;\;\;\;\;\;\;\;\;\;\; \nonumber \\
  1+(z-1)^2[ F_2(1-F_0)+F_1^2+F_0-2F_1 ] \;,
\label{eq:criterion}
\end{eqnarray}


\end{document}